# Tunguska similar impacts and origin of life

Andrei E. Zlobin

*Vernadsky State Geological Museum, Russian Academy of Sciences*
*Mokhovaya 11/11, 125009, Moscow, Russian Federation*
*e-mail: z-tunguska@yandex.ru*

*Abstract.* The author suggests new vision of mechanism of initiation of life on the planets after Tunguska similar impacts. This mechanism takes into consideration not only incoming cosmic organic substance but also information, which is connected to this substance. Mathematical metrics of atom of hydrogen is deduced which may be used for pattern recognition algorithm. In accordance to author's opinion, similar algorithm can promote evolution (transformation) of inert organic substance into living substance. The fact of a survival of vegetation after the Tunguska event is analyzed especially. Also the author checked up his probably Tunguska meteorites by strong magnet. The presence of magnetic substance was detected with concentration of $10^{-2}$ % during this test.

## 1. Introduction

It is well known that academician V.I.Vernadsky demonstrated great interest to conditions of initiation of life on the Earth [Vernadsky, 1931]. Close correspondence between cosmic phenomena and initiation of life on the planet was mentioned by Vernadsky. Also Vernadsky described several functions of biosphere from position of chemical phenomena. Academician V.G.Fesenkov concluded that Tunguska event was caused by comet impact, and he mentioned that comets are able to disseminate such elements as H, C, N, O [Fesenkov, 1964]. Ideas of interaction between cosmos and substance of the Earth were supported by A.L.Chizhevsky, who considered the life as more cosmic than terrestrial phenomenon [Chizhevsky, 1995]. The origin of life also was interesting for academician N.V.Vasilyev, who discussed this theme in connection to Tunguska event [Vasilyev, 2004]. Academician of Russian Academy of Medical Sciences N.V.Vasilyev was coordinating investigation of Tunguska phenomenon during 40 years and he noted considerably influence of the Tunguska impact on ecology. Moreover, N.V.Vasilyev described a number of biological consequences of the Tunguska explosion. For example, too quick growth of trees after Tunguska impact was mentioned by Vasilyev. A lot of other authors are discussing significance of cosmic phenomena, comets and meteorites for initiation of life too. The author of this paper has the opinion that one more aspect of origin of life must be taken into consideration. This aspect is connected not only to substance, but also to information, which is necessary for initiation of life. That is why, the incoming cosmic substance of comets and meteorites should be analyzed from position of incoming of additional information too.

## 2. Comets and substance of the Tunguska body

Let's pay attention to intensive investigation of comets with the help of space missions. Results of this activity gave us a lot of new information about different comets. Close investigation of several comets was carried out during last decades. For example, Halley comet [Sagdeev et al., 1986, 1988], [Keller et al., 1986] and Tempel 1 comet [A'Hearn et al., 2005] should be mentioned. Also "Deep Impact" spacecraft investigated 103P/Hartley 2 comet. Earth ocean-like water in 103P/Hartley 2 comet was noted [Hartogh, P. et al., 2011]. Organics was found in comet 81P/Wild 2 due to "Stardust" mission and caution must be taken in interpreting measurements of organics in "Stardust" samples [Sandford and Brownlee, 2007]. Some important parameters of Tunguska comet were determined during mathematical quasi three-dimensional modeling, including water ice as main substance of mass of the comet [Zlobin, 2007]. There is good correspondence between results of this mathematical modeling and data of missions. For example, the form of Tunguska comet [Zlobin, 2007] is considerably similar to the form of 103P/Hartley 2 comet. Water ice is mentioned in most of cases as considerable component of comet's nucleus. The density of Tunguska comet was obtained during mathematical modeling as 0.6 g/cm$^3$ and this value is in good correspondence to Halley comet etc. If to assume compound structure of the Tunguska comet's nucleus, then seems possible to discuss presence of stony bodies into the comet's ice. After destruction of comet's nucleus in atmosphere of the Earth this bodies was able to fall on the ground as meteorites. In particular, D.F.Anfinogenov and L.I.Budaeva are not excludes possibility of discovering of large fragments of Tunguska cosmic body [Vasilyev, 2004]. As illustration of this opinion, in 1988 three probable Tunguska meteorites were already found at the bottom of Khushmo River's shoal (Fig. 1). Cautious investigation continues of these stones with the help of different methods and without destruction of samples [Zlobin, 2013].

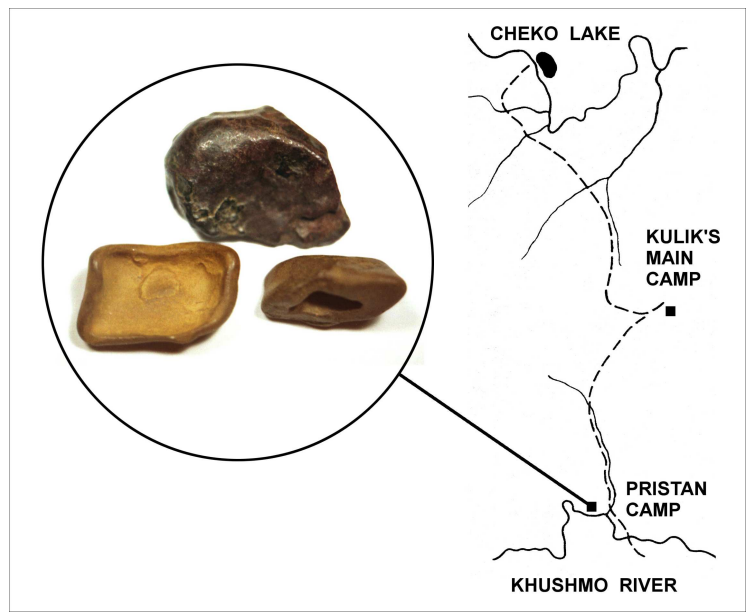

Fig. 1. Fragment of schematic map of the Tunguska region is presented on this figure. The place is shown where three probably Tunguska meteorites were found by the author in 1988.

It is necessary to remind that for the first time considerably fragment of melted glass-like substance was already discovered by L.I.Kulik [Kulik, 1939]. There is some information that traces of Ni were found by Kulik too. A lot of small magnetic spheres were collected during meteorite complex expedition in 1961 [Florensky, 1963]. Increased presence of Ni was noted by K.P.Florensky as the proof of cosmic origin of magnetic spheres. S.P.Golenetsky and others published their results concerning analysis of layers of peat-bogs [Golenetsky et al., 1977]. Many chemical elements were discovered during this analysis; moreover, Fe and Ni concentration was noted as increased too. Authors concluded that results of analysis correlates to substance of comets. Also Fe and Ni were discovered in particles from resin of trees [Longo, 1996] and this fact was in good correspondence to data of Golenetsky and others. Interesting results was obtained during investigation of small spherical particles with gas blebs [Dolgov, 1980]. Presence of hydrogen was determined in gas blebs and air component was absent. These spherules were found in the Tunguska region and researchers made the conclusion concerning comet origin of the substance. E.M.Kolesnikov and others discovered anomaly in isotopic composition of hydrogen in the peat from the place of explosion of Tunguska cosmic body [Kolesnikov et al., 1995]. The same opinion was presented in this study concerning comet origin of Tunguska body.

In accordance to author's opinion the Tunguska cosmic body really has more chances to be recognized as a comet but not an asteroid. There is not any large crater. Also large quantity of cosmic mineral substance was not found. Now we have only three melted stones which have to be further examined as probably Tunguska meteorites. If to discuss connection between comets and origin of life, initially it is necessary to check up version of presence in these stones any magnetic markers of cosmic substance. It is useful to present arguments concerning possibility of survival of organic substance during comet impact. Finally, some informational model should be suggested, which can demonstrate possibility of self-recognition among atoms of cosmic organic substance. Such algorithm of self-recognition seems necessary for further modification of simple organic substance into simple forms of life.

### 3. Test of probably Tunguska meteorites by strong magnet

Description of special magnetic test is presented below, which base on usage of strong magnet. This test was carried out for the purpose of determination of any magnetic substance in the volume of every melted stone from Khushmo River's shoal. All three stones were investigated by magnetic test: "whale", "dental crown" and "boat". The view of stones in comparison with the magnet is presented on Fig. 2. The size of magnet was 79 x 56 x 24 mm. Cargo characteristic of the magnet can be estimated with the help of simple experiment. This magnet can attract iron cargo of 1.5 kg from the distance of 1 cm. Let us repeat the weight of stones: "whale" - 10.4 g, "dental crown" - 1.6 g and "boat" - 2.3 g. The author used two methods of magnetic test. In accordance to first method, the influence of magnet on stones was primary estimated by simple moving up of magnet to every stone. All attempts were negative. It was not registered any motion of stones in direction to magnet. Second method of testing was considerably more sensitive when the author decided to use precise laboratory weighing machine (scales). Investigated stone was hanged on the lever of weighing machine with the help of long cotton string (1 meter). Another lever of weighing machine (without long string) was used for correction weight. Before tests, left and right lever of weighing machine were strictly balanced. Then the magnet was moving under stone and fluctuations of levers were registered by pointer (arrow) of weighing machine. It was good visible that every stone was slowly and weakly attracted to strong magnet. Degree of magnetic influence was approximately equal for "dental crown" and "boat", but little more for "whale" stone.

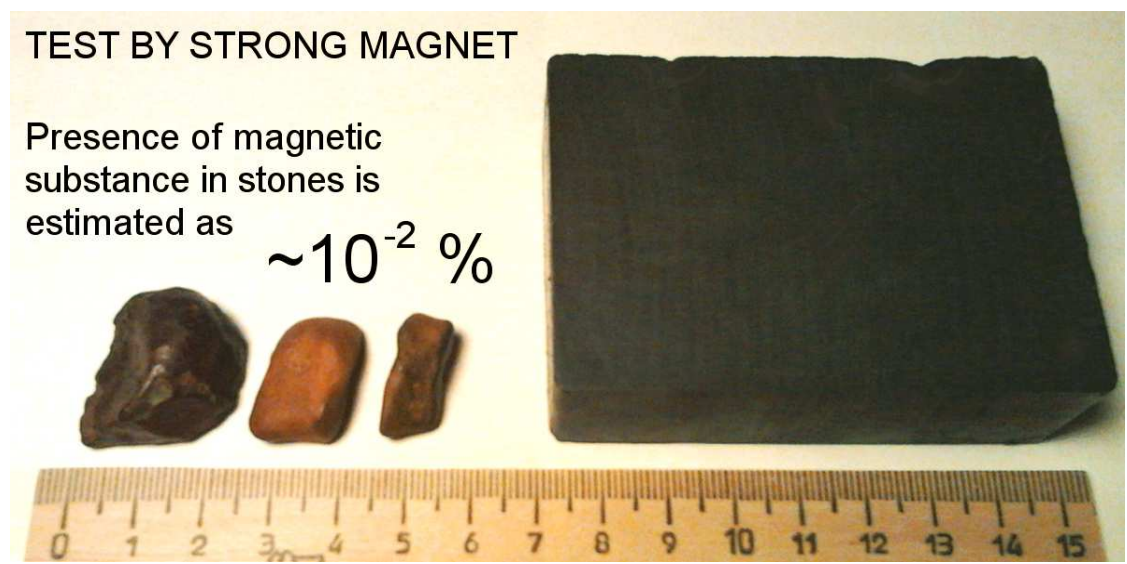

Fig. 2. Tested stones "whale", "dental crown" and "boat" in comparison with strong magnet and ruler

Several other stones from author's Khushmo collection were tested with the same magnetic test. There were not registered any fluctuations of levers of weighing machine during motion of strong magnet under stones. This became one more confirmation that three probably meteorites considerably differ from another stones, which were collected at the bottom of Khushmo River's shoal. Also successful magnetic test demonstrated that three melted stones probably could be fragments of single parent body.

The author made the attempt to determine concentration of magnetic substance in "whale" stone. He used for this purpose special "test stone" from Khushmo collection without traces of melting and without magnetic properties. This stone has approximately the same weight as "whale" stone. Little iron particle was attached to "test stone" to achieve the same fluctuations of weighing machine as at magnetic test in case of "whale" stone. This method gave first estimation of concentration of magnetic substance in "whale" stone as ~ $10^{-2}$ % (crude guess). The author has the opinion that iron [Fe] is possible magnetic substance in three melted stones. It seems this opinion may be confirmed with rusty color of melted stones. Certainly further investigation with more accurate chemical methods has to be carried out.

## 4. Survival of vegetation after the Tunguska impact

Survival of vegetation during the Tunguska impact can be taken into consideration in case of analysis of survival of any organic substance. Heat impulse of Tunguska impact was already determined with good methods of proof. The author of this paper obtained values 13 - 30 $J/cm^2$ for the level on the ground [Zlobin, 2007] and values 280 and 420 $J/cm^2$ in considerably altitude above the ground [Zlobin, 1996, 1997, 2007, 2013]. Equation of thermal balance was used in last case for micro-meteorites [Whipple, 1950, 1951]. Approximately the same range of values was mentioned by experts of Sukachev Institute of Forest, University of Arizona and Institute of Monitoring of Climatic and Ecological Systems [Vaganov et al., 2004]. These authors concluded that unlikely heat impulse exceeded 300 $J/cm^2$ (no signs of tree's crown fire) and the minimum heat impulse was estimated by these authors as 25 $J/cm^2$.

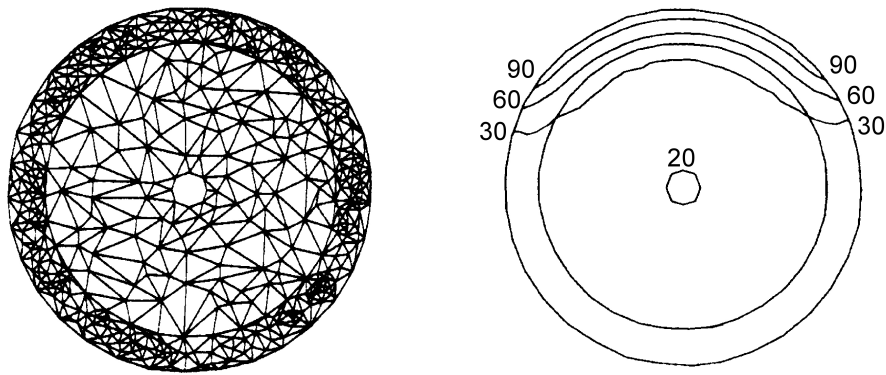

Fig. 3 FEM grid with random generated nodes and temperature distribution in cross section of branch of pine (Celsius degrees). Diameter of branch - 10.5 mm. Thickness of rind - 1 mm. Heat impulse - 13 $J/cm^2$. Initial temperature - 20 degrees. Duration - 2 seconds. [Zlobin, 1996].

It is well known the fact of survival of trees at central region of the Tunguska impact [Vasilyev, 2004]. A lot of such living trees has new crown, which was formed again after influence of shock wave and burn of branches. The author made calculations of temperature distribution in cross section of branches during heating of tree by thermal radiation of Tunguska fire-ball [Zlobin, 1996]. Certainly, the author took into consideration presence of early-dew in all calculations too. Two-dimensional finite element method was used for these calculations and typical unsteady temperature field is presented in Fig.3. There is good visible heat influence near the external surface, which makes thermal damage in sectorial region of branch (in all depth of rind). However, any heat influence is absent in another regions of branch. That is why survival of many branches was provided too.

Also the author made calculations of temperature distribution in the depth of ground [Zlobin, 1997]. It was shown that some influence of thermal radiation of Tunguska explosion penetrates into the ground only to the depth of 1 - 2 mm. This result is in good correspondence to the fact of survival of seeds of trees in the ground in central region of the impact [Nekrasov et al., 1967]. During expedition of 1988 the author made more than ten prospect-holes in the Sphagnum fuskum peat-bogs at different places of central region of the Tunguska impact. The layer of fire of 1908 was detected accurately, and this layer was close inspected. There were good visible small parts of burned thin branches and roots of vegetation, but some of samples not looked like burned completely. It is one more confirmation that heat impulse was not considerably more than 13 - 30 $J/cm^2$. In 1988 the author inspected many peat-bogs around the center of Tunguska catastrophe and it was good visible that growth of vegetation was intensive during many decades after the event (Fig. 4).

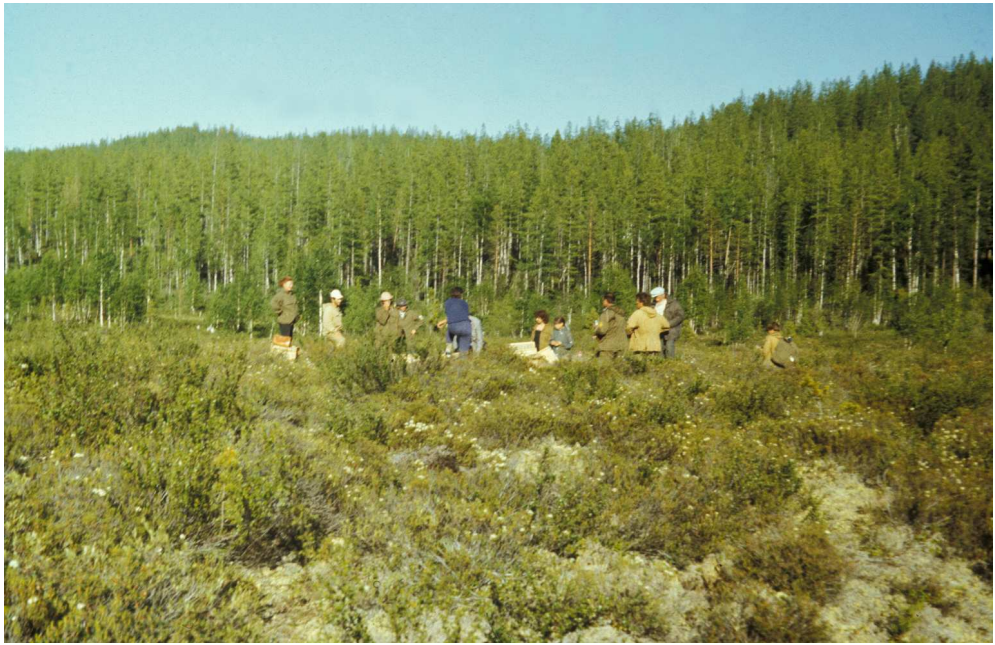

Fig. 4 View of vegetation near Kulik's main camp. Participants of 1988 expedition among hummocks of peat-bog and near new forest [Zlobin Photo, 1988 expedition]

E.A.Vaganov and others noted one more interesting fact concerning damages of trees. In accordance to disrupted tracheids in the 1908 tree's ring, authors noted, that mechanical stress needed to cause the deformation of differentiating tracheids seen in the trees close to the epicenter is greater than needed to fell trees. Authors mentioned the assumption of complicated pattern of shock waves, caused with interaction of blast wave front with local topographic elements [Vaganov et al., 2004]. Also it seems useful to remind concerning complicated pattern of shock waves, caused with several powerful explosions during falling of fragments of Tunguska cosmic body [Zlobin, 2007]. As the author demonstrated earlier, pattern of the field of heat influence indicates four local flashes. Different views of forest-fall were good visible in 1988, and many trees were broken by high energy influence (Fig. 5). However, survival of many trees was provided and growth of new forest became the best illustration of restoring of biosphere.

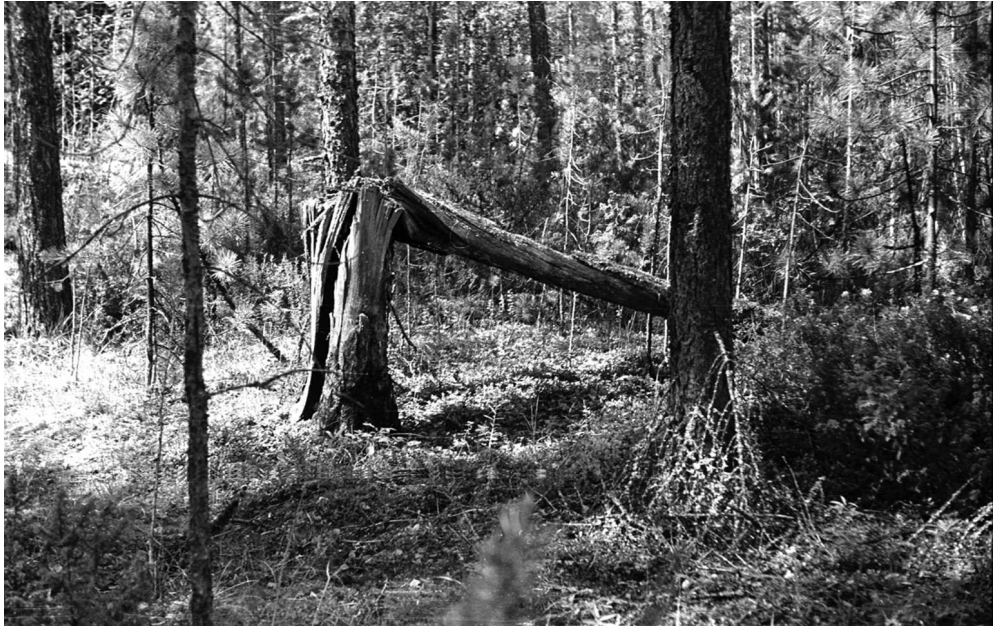

Fig. 5 Central region of the Tunguska impact. This tree was broken by high energy influence of Tunguska cosmic body. Trees of new forest are visible around [Zlobin Photo, 1988 expedition].

We can make conclusion that in conditions of high temperatures and powerful shock waves of Tunguska impact, organic substance on the surface of the Earth can survive. There are more high temperature and pressure in the volume of fire-ball. However, in case of comet with compound structure, considerable mass of low temperature ice is able to protect internal stony bodies from premature heating and breaking during motion through the atmosphere. Moreover, thin crack was discovered last time when the "whale" stone from Khusmo River's shoal was investigated with high resolution magnifier (confirmation of considerably stresses [Melosh, 1989]). Similar cracks can include comet ice and simple organic substance too. This substance can fall on the surface of the Earth with meteorites. Therefore, we can not exclude possibility of incoming of cosmic organic substance with comets and meteorites. Certainly, it is possible to imagine that organic life was delivered into the Earth from other space bodies. However, there is the question: how organic life was initiated on these other bodies?

V.I.Vernadsky had the opinion that the life appeared considerably quickly, and many different kinds of life were initiated simultaneously. To his mind, this conclusion can explain difference between geological layers of inert substance (inert minerals) and later geological layers with numerous traces of life. The velocity of distribution of life was mentioned by Vernsdsky as 1000 -10000 centimeters per second, and all surface of planet could be covered by life during several days. In accordance to Vernadsky the process of evolution of life began after this [Vernadsky, 1931]. The logic, mentioned above, makes possible to explain initiation of life by injection of some special informative substance into the Earth. One more explanation may be connected to achievement of necessary climatic conditions for this informative substance. The author of this paper focused his attention on hydrogen as the most known and widespread substance in the universe [Zlobin, 1996, 2010]. Some ideas concerning informational properties of hydrogen the author try to illustrate below.

## 5. Mathematical metrics of atom of hydrogen

V.I.Vernadsky mentioned some aspects which are connected to problem of initiation of life and biosphere on the Earth [Vernadsky, 1931]. Among these aspects are ocean water, gas functions, pressure and temperature, climate etc. However three most interesting ideas were mentioned by Vernadsky especially. The first idea is known for a long time: "omne vivum e vivo" (it means that every living thing descends from living thing). The second idea means that all living things do not have strict symmetry, and left and right side of every living thing are different. Thus, all living things are characterized with the property of asymmetry. The third idea is that this asymmetry may be described mathematically as infringement of symmetry.

The author found mathematical expression, which seems useful for further theoretical analysis of the phenomenon of life [Zlobin, 1996, 2010].

Initially let's remind the view of Fibonacci row of numbers [Vorobyov, 1969]

$$1, 1, 2, 3, 5, 8, 13, 21, 34, 55, 89, 144, 233, 377\ldots$$

where values of a series are calculated with recurrent expression ($n>2$)

$$u_n = u_{n-1} + u_{n-2}$$

also let's take two expressions

$$\frac{u_{n+1}}{u_n} \quad \text{and} \quad \frac{1}{\left(1+\frac{1}{u_n}\right)^{u_{n+1}}}$$

where last expression we can transform to

$$\frac{1}{\left(1+\frac{1}{u_n}\right)^{u_{n+1}}} = \frac{1}{\left(\left(1+\frac{1}{u_n}\right)^{u_n}\right)^{\frac{u_{n+1}}{u_n}}}$$

let's $\dfrac{u_{n+1}}{u_n}$ multiply by $\dfrac{1}{\left(\left(1+\dfrac{1}{u_n}\right)^{u_n}\right)^{\frac{u_{n+1}}{u_n}}}$

then the result will be

$$\frac{u_{n+1}}{u_n} \cdot \frac{1}{\left(\left(1+\frac{1}{u_n}\right)^{u_n}\right)^{\frac{u_{n+1}}{u_n}}} = \frac{\frac{u_{n+1}}{u_n}}{\left(\left(1+\frac{1}{u_n}\right)^{u_n}\right)^{\frac{u_{n+1}}{u_n}}}$$

if $n \to \infty$ then we can write

$$\lim_{n \to \infty} \frac{u_{n+1}}{u_n} = \Phi$$

and

$$\lim_{n \to \infty}\left(1+\frac{1}{n}\right)^n = e$$

where

$\Phi$ – "extreme and mean ratio" or "golden ratio"
$e$ – Napier number

then we have simple expression

$$\frac{\Phi}{e^{\Phi}}$$

the property of this expression is that it approximately equal to $\frac{1}{\pi}$

$$\frac{\Phi}{e^{\Phi}} \approx \frac{1}{\pi}$$

where

$\pi$ – Ludolph number

the equality will be very accurate if $\frac{1}{\pi}$ to multiply by special factor

let's consider this factor as $j=1.0079\ldots$ (irrational one number)

thus

$$\frac{\Phi}{e^\Phi} = \frac{j}{\pi}$$

or finally

$$\frac{\pi \cdot \Phi}{e^\Phi} = j \qquad (1)$$

certainly it mean, that

| $\frac{\pi \cdot \Phi}{e^\Phi} = j$ | $j \cdot \left(\frac{e^\Phi}{\pi}\right) = \Phi$ | $\left(\frac{\pi}{j \cdot \varphi}\right)^\varphi = e$ | $j \cdot \left(\frac{e^\Phi}{\Phi}\right) = \pi$ |

where

| $j = 1.0079...$ | $\Phi = 1.6180...$ | $e = 2.7182...$ | $\pi = 3.1415...$ |

and $\varphi = \frac{1}{\Phi}$

The most wonderful property of obtained mathematical expression (1) is that the factor $j=1.0079...$ simultaneously coincides to the value of atomic mass $m_H$ of hydrogen with high accuracy. Let's demonstrate that this expression may be used as system of measurements (metrics) for analysis of atoms. It seems very convenient, because all values during such analysis will be relative.

For example, below we make analysis for the atom of hydrogen (H). There is good visible correspondence between mathematical expression (1) and the length of the circle. Let's consider the value $j=1.0079...$ as the length of the circle of atom of hydrogen

$$L_H = \frac{\pi \cdot \Phi}{e^\Phi} = j$$

where index "H" relates to all values obtained during analysis of atom of hydrogen.

Then we can write expressions for radius and diameter of the atom

$$R_H = \frac{\Phi}{2 \cdot e^\Phi} = \frac{j}{2\pi} \qquad D_H = \frac{\Phi}{e^\Phi} = \frac{j}{\pi}$$

Expressions for area of sphere and volume of sphere

$$S_H = 4\pi R^2 = 4\pi \left(\frac{\Phi}{2 \cdot e^\Phi}\right)^2 = 4\pi \cdot \frac{1}{4} \cdot \frac{j^2}{\pi^2} = \frac{j^2}{\pi}$$

$$V_H = \frac{4}{3}\pi R^3 = \frac{4}{3}\pi \left(\frac{\Phi}{2 \cdot e^\Phi}\right)^3 = \frac{4}{3}\pi \cdot \frac{1}{8} \cdot \frac{j^3}{\pi^3} = \frac{j^3}{6\pi^2}$$

If atomic mass $m_H \approx j$ then average density of atom of hydrogen

$$\rho_H = \frac{j}{V_H} = \frac{j}{\frac{j^3}{6\pi^2}} = \frac{6\pi^2}{j^2}$$

Let's include all obtained values into single Table 1. Approximate values will be included into the last column too (if to take into consideration approximation $j \approx 1$)

Table 1

| Value | More accurate | Approximate |
|---|---|---|
|  |  |  |
| $m_H$ | $j$ | $\approx 1$ |
| $L_H$ | $j$ | $\approx 1$ |
| $R_H$ | $\dfrac{j}{2\pi}$ | $\approx \dfrac{1}{2\pi}$ |
| $D_H$ | $\dfrac{j}{\pi}$ | $\approx \dfrac{1}{\pi}$ |
| $S_H$ | $\dfrac{j^2}{\pi}$ | $\approx \dfrac{1}{\pi}$ |
| $V_H$ | $\dfrac{j^3}{6\pi^2}$ | $\approx \dfrac{1}{6\pi^2}$ |
| $\rho_H$ | $\dfrac{6\pi^2}{j^2}$ | $\approx 6\pi^2$ |

Interesting results can be obtained if to analyze some another atoms with the help of the metrics. For example: when using this relative system of measurements, diameter of atom of hydrogen (H) approximately equal $\approx 1/\pi$, diameter of atom of gold (Au) approximately equal $\approx 1$, and diameter of atom of potassium (K) approximately equal $\approx \Phi$ (golden ratio). This wonderful fact really confirms correctness of the term "golden ratio" as historical tradition. Let's demonstrate this conclusion with concrete calculations. Size of every of mentioned atom is presented in Table 2 [Glinka, 1979].

Table 2

| Designation | Chemical element | Radius of atom, nm |
|---|---|---|
|  |  |  |
| H | hydrogen | 0.046 |
| K | potassium | 0.236 |
| Au | gold | 0.144 |

The length of circle of hydrogen's atom will be used as the unit.

$$L_H^{nm} = 2\pi R_H^{nm} = 2\pi \cdot 0.046 = 0.289 nm \quad \text{(unit)}$$

$$D_K^{nm} = 2R_K^{nm} = 2 \cdot 0.236 = 0.472 nm$$

$$D_{Au}^{nm} = 2R_{Au}^{nm} = 2 \cdot 0.144 = 0.288 nm$$

Then

$$D_K = D_K^{nm} / L_H^{nm} = 0.472 / 0.289 = 1.63$$

where $\Phi \approx 1.62$ and the difference between 1.63 and 1.62 is $\approx 1\%$

$$D_{Au} = D_{Au}^{nm} / L_H^{nm} = 0.288 / 0.289 = 0.997 \approx 1$$

where the difference between 0.997 and 1 less than 1%

and finally

$$D_K / D_{Au} \approx \Phi \quad \text{(golden ratio)}$$

The author of this paper makes note that potassium is very important chemical element in biology. Also it seems interesting that Fibonacci numbers and golden ratio are close connected to phenomena of life. For the first time Fibonacci numbers were mentioned in the task of rabbits born [Vorobyov, 1969]. If to take into consideration the paper by Vernadsky [Vernadsky, 1931], we can note good correspondence between obtained metrics and Vernadsky's ideas concerning origin of life.

We strongly demonstrated that metrics of atom of hydrogen connects between each other the idea of substance, idea of form and idea of number.

It seems, there is good visible reflection of the idea "omne vivum e vivo"

$$\Phi = j \cdot \left( \frac{e^{\Phi}}{\pi} \right)$$

The idea of asymmetry is good visible too

$$\frac{\pi \cdot \Phi}{e^{\Phi}} = 1.0079...$$

and there is not any doubt, that the infringement of symmetry is expressed mathematically here.

Due to obtained metrics (1) interaction between atoms may be presented as simple analogue of thinking. Pattern recognition algorithm here is possible which based on strongly mathematical laws. By another words, we can imagine initiation of life not only as evolution of inert organic substance, but as evolution of "algorithmically thinking" substance. Also we have to remember the fact of deduction of the metrics from Fibonacci row of numbers. This may be theoretical hint concerning complicated internal logical structure of the atom too.

Now let's return to the Tunguska event and remind concerning discovering of hydrogen as probably substance of the cosmic body. If the Tunguska body really was a comet then considerably mass of water and hydrogen was delivered to the Earth. The same substance could be delivered with similar impacts during prehistoric times. Simultaneously the information was delivered to the Earth, which was mentioned above as simple "algorithms of thinking". The author considers that the presence of additional information (metrics for pattern recognition algorithm) became the condition of initiation of life in presence of simple inert organic substance too. Also the Tunguska event seems confirms that process of forming of the Earth and initiation of life continues till now.

If to follow this mind the initiation of life on another planets is possible too. Moreover, we must remember concerning other necessary conditions of life (temperature, pressure etc.). It is important to notice, that the life on all planets must be considerably similar. This conclusion follows strongly from mathematical laws (mathematical algorithms), when similar metrics is in usage and similar organic substances are present. Once again, hydrogen is the most widespread substance in the universe. That is why the metrics of atom of hydrogen is the most widespread too.

# 6. Conclusion

After 25 years of research the author prefers comet nature of the Tunguska impact. This event is accompanied with massive incoming of cosmic water on the Earth and the same incoming of cosmic hydrogen. The test by strong magnet was carried out with three probable Tunguska meteorites from Khushmo River's shoal. The test was successful and the presence was shown of magnetic substance in every stone ($\sim 10^{-2}$ %). Mathematical modeling of heat processes confirms possibility of survival of vegetation during comet impact. The crack was discovered in the body of "whale" stone and the author consider that survival of simple organic cosmic substance is possible in such cracks too. Also the author obtained mathematical expression which can be used as metrics during analytical studies of atoms. Atom of hydrogen (H) is selected as the base of this metrics. It is suggested that initiation of life could be realized by incoming of simple cosmic organic substance simultaneously with incoming of additional information in the form of mentioned metrics of hydrogen. Pattern recognition algorithm is possible on the base of the metrics and this algorithm could be realized during evolution from inert to living substance.

## Acknowledgments

I very thankful to all people who were supporting my Tunguska study during three decades. Also I very thankful to administration and my colleagues in Vernadsky State Geological Museum (RAS) on possibility to work with actual information concerning scientific activity of V.I.Vernadsky and L.A.Kulik.